\documentclass[aps,pra,twocolumn,groupedaddress,showpacs]{revtex4}
\usepackage{mathrsfs}

\usepackage{amsmath}
\usepackage[toc,page,title,titletoc,header]{appendix}

\begin{document}
\title{Quantum form of Nonlinear Maxwell equations}
\author{Ying Chen}
\author{Fu-Lin Zhang}
\affiliation{Theoretical Physics Division, Chern Institute of
Mathematics, Nankai University, Tianjin 300071, P.R.China}
\author{Jing-Ling Chen}
\email[Email:]{chenjl@nankai.edu.cn} \affiliation{Theoretical
Physics Division, Chern Institute of Mathematics, Nankai University,
Tianjin 300071, P.R.China}
\date{\today}

\begin{abstract}
We generalize Maxwell equations  which describe the vacuum of
quantum electrodynamics into the quantum form. This nontraditional
approach is different from the widely used theory----Quantum
Electrodynamics. From another viewpoint, it could be a direction for
interpreting quantum theories properly.
\end{abstract}
 \maketitle

\section{Introduction \label{intro}}
%In principle, quantum electrodynamics could solute all
%electromagnetic field problems,but it should not be an obstacle to
%study electromagnetic field within the Quantum Mechanics especially
%when it will bring some interesting and thought-provoking results.
%First quantized quantum equations are one particle equations which
%can describe also many non-interacting particles due to the
%statistical features of the equations.(1, 2)

Since quantum electrodynamics is quite mature, it is not widely
known that Maxwell equations are first quantized quantum equations
of a photon after the Planck constant $\hbar$ was multiplied to it
\cite{Ger3}. In fact, there has been an almost generally accepted
way to change free noninteracting Maxwell equations into the form of
Dirac equation\cite{Greiner1}. The main purpose of this letter is to
 study Maxwell equations with self-interaction of a photon.

 Much attention was paid to study spinor formulation of Maxwell
equations after Dirac \cite{Dirac} find the relativistic equation
for a particle with spin-$1/2$. Landau and Peierls photon wave
functions \cite{Landau} could ensure that the modulus squared of
wave function be the probability density of a photon, but this
approach has some weaknesses as Pauli \cite{Pauli} had pointed out.
And Laporte
%and Uhlenbeck\cite{L&U}gave Maxwell's equations' spinor formulation
%with a complex 3-vector wave function.
%\begin{eqnarray}\label{Sch}
%g_{\dot{1}\dot{1}}=2(\Psi2+i\Psi_1)\\
%g_{\dot{2}\dot{2}}=2(\Psi_2-i\Psi_1)\\
%g_{\dot{1}\dot{2}}=g_{\dot{2}\dot{1}}=2\psi_3,
%\end{eqnarray}
%where $\psi_j=E_j+iB_j$.This work was enlightened by the complex
and Uhlenbeck \cite{L&U} gave Maxwell equations' spinor formulation
\begin{eqnarray}\label{Sch} \nonumber
&&g_{\dot{1}\dot{1}}=2(\Psi_2+i\Psi_1),\\ \nonumber
&&g_{\dot{2}\dot{2}}=2(\Psi_2-i\Psi_1),\\ \nonumber
&&g_{\dot{1}\dot{2}}=g_{\dot{2}\dot{1}}=2\Psi_3, \nonumber
\end{eqnarray}
$g_{ik}$ is a spinor of second rank and $\psi_j(\psi_j=E_j+iB_j)$ is
a complex 3-vector wave function whose real and virtual parts are
electric and magnetic fields. The most crucial step was made by
Silberstein \cite{Sil} who proved the possibility to describe
Maxwell equations in terms of a 3-vector entities. However, it is
Oppenheimer \cite{Opp} who noticed quantum nature of Maxwell
equations for the first time(there is other opinion about this
\cite{1974}). And Good \cite{Good} gave more clear interpretation
about the similarity between Maxwell equations and Dirac equation,
the new Maxwell equations transformation properties and conservation
theorems. Following the idea of
Riemann-Silberstein-Majorana-Oppenheimer
 approach, Bialynicki-Birula \cite{IBB1,I2,I3} gave the more detailed
consideration about single photon quantum mechanics in resent years.

In addition, Gersten \cite{Ger1,Ger2,Ger3} finds an alternative
approach to gain the Dirac form of the Maxwell equations, which
starts from the first principle and the derivation process is just
like Dirac equation. If the momentum $\mathbf{p}$ is written in form
of 3-vectors column
\begin{eqnarray}\nonumber
\mathbf{p}= \begin{bmatrix}p_1\\p_2\\p_3\end{bmatrix},
&&\mathbf{p}^\texttt{T}= \begin{bmatrix}p_1&p_2&p_3\end{bmatrix}.
\end{eqnarray}
Decomposing the left side of the relationship between energy and
momentum of electrodynamic field after multiply a $3\times 3$ unit
matrix $\texttt{I}$, we obtain
\begin{eqnarray}\nonumber
\biggr(\frac{E^2}{c^2}-\mathbf{p}^\texttt{T}\mathbf{p}\biggr)\texttt{I}=
\biggr(\frac{E}{c}\texttt{I}-\mathbf{p}^\texttt{T}
\mathbf{s}\biggr)\biggr(\frac{E}{c}\texttt{I}+\mathbf{p}^\texttt{T}
\mathbf{s}\biggr) - \mathbf{p}\mathbf{p}^\texttt{T} =0,
%-\begin{bmatrix}p_x^2&p_xp_y&p_xp_z\\\nonumber
%p_yp_z&p_y^2&p_yp_z\\\nonumber p_zp_x&p_zp_y&p^2_z\end{bmatrix}
\end{eqnarray}
where $\mathbf{s}=[s_1, s_2 ,s_3]^{\texttt{T}}$ is a spin-1 vector
matrix with three components
\begin{eqnarray}
s_1=\begin{bmatrix}0&0&0\\
0&0&-i\\
0&i&0\end{bmatrix},s_2=\begin{bmatrix}0&0&i\\
0&0&0\\
-i&0&0\end{bmatrix},
s_3=\begin{bmatrix}0&-i&0\\
i&0&0\\
0&0&0\end{bmatrix}.
\end{eqnarray}
Then the photon equation is obtained,
\begin{eqnarray}\label{111}
\biggr(\frac{E^2}{c^2}-\mathbf{p}^\texttt{T}\mathbf{p}\biggr)\psi
=\biggr(\frac{E}{c}\texttt{I}-\mathbf{p}^\texttt{T}
\mathbf{s}\biggr)\biggr(\frac{E}{c}\texttt{I}+\mathbf{p}^\texttt{T}
\mathbf{s}\biggr)\psi -\mathbf{p}\mathbf{p}^\texttt{T}\psi =0,
\end{eqnarray}
where $\psi$ is a 3 components wave function. At last, Eq.
(\ref{111}) holds for the following two expressions are satisfied
\begin{eqnarray}
&&\big(\frac{E}{c}\texttt{I}+\mathbf{p}^\texttt{T}\cdot \mathbf{s}\big)\psi=0,\\
&&\mathbf{p}^\texttt{T}\cdot \psi =0.
\end{eqnarray}
If the quantum operator substitutions and the wave function
substitution are made as follows
\begin{eqnarray}
&&E\Rightarrow i\hbar \frac{\partial}{\partial
t},\mathbf{p}\Rightarrow -i\hbar
\mathbf{\bigtriangledown},\\
&&\psi=\mathbf{E}-i \mathbf{B},
\end{eqnarray}
the classical Maxwell equations could be obtained
\begin{eqnarray}
i\hbar  \frac{\partial\psi}{\partial t}=-c  \mathbf{s}\cdot
\mathbf{p} \psi.
\end{eqnarray}
And  the Gersten's approach return to Bialynicki-Birula's.

In addition, there are many works
\cite{1956,19,82,1985,1988,1994,1995,1998,I05,I052,06,2009}
concerning the Dirac-like form of Maxwell equations. In this paper,
we extend linear quantum theory of Maxwell equations to nonlinear
situation caused by the possibility of creating virtual particles in
vacuum. Starting from Lagrange with an corrections in higher orders
in E and B, we obtained the nonlinear Maxwell equations, and then
try to convert it into the nonlinear Schr$\ddot{o}$dinger equation
with the same approach and wave function as linear situation.

\section{Nonlinear Schr$\ddot{O}$dinger form of Nonlinear Maxwell equations\label{intro}}
The Lagrangian of Quantum electrodynamics in vacuum was given by
Heisenberg and Euler\cite{Greiner}. It can describe the phenomenon
of optical birefringence and experiments of measuring the vacuum
birefringence.

In the limiting case of the stationary and homogeneous
electromagnetic field, the exact expression of $\mathscr{L}$ would
be
\begin{eqnarray}\label{000}
\mathscr{L}=-\frac{1}{4}(F_{\mu\nu}F^{\mu\nu})+\frac{1}{8}k_1(F_{\mu\nu}F^{\mu\nu})^2+\frac{1}{8}k_2(F_{\mu\nu}\tilde{F}^{\mu\nu})^2,
\end{eqnarray}
where $F_{\mu\nu}$ is the electromagnetic field tensor, and
$\tilde{F}^{\mu\nu}$ is  the dual-field tensor contracted by
$F_{\mu\nu}$ with completely antisymetric unit tensor (the
Levi-Civita tensor) $\epsilon^{\mu\nu\gamma\delta}$. $F_{\mu\nu}$
has the relation with $A_\mu$ as follow
\begin{eqnarray}
F_{\mu\nu}=\partial_\mu A_\nu -\partial _\nu A_\mu,
\end{eqnarray}
$A_\mu$ is vector potential in the electromagnetic field. so
\begin{eqnarray}
&&F_{\mu\nu}=\begin{bmatrix}0&-E_1&-E_2&-E_3\\
E_1&0&B_3&-B_2\\
E_2&-B_3&0&B_1\\
E_3&B_2&-B_1&0\end{bmatrix},\\
&&\tilde{F}^{\mu\nu}=\frac{1}{2}\epsilon^{\mu\nu\gamma\delta}F_{\gamma\delta}
=\begin{bmatrix}0&B_1&B_2&B_3\\
-B_1&0&-E_3&E_2\\
-B_2&E_3&0&-E_1\\
-B_3&-E_2&E_1&0\end{bmatrix}.
\end{eqnarray}
For notation here, Greek indices $\mu$, $\nu$,$\gamma$ and $\delta$
run from 0 to 3.

And $k_1$, $k_2$ determine the magnitude of the nonlinear correction
\begin{eqnarray}
k_1=\frac{4\alpha^2}{45m^4_e},&&  k_2=\frac{7\alpha^2}{45m^4_e},
\end{eqnarray}
$\alpha$ is fine-sructure constant: $\alpha=\frac{e^2}{4\pi\hbar
c}\approx \frac{1}{137}$.
%\begin{eqnarray}
%\alpha=\frac{e^2}{4\pi\hbar c}\approx \frac{1}{137}
%\end{eqnarray}

The last two terms of Eq. (\ref{000}) are the corrections in
$B^2-E^2$ and $B\cdot E$ which are Lorentz invariants, so
$\mathscr{L}$ is invariable in any frame of reference. $\mathscr{L}$
satisfies the following Lagrange's equation
\begin{eqnarray}
\frac{\delta \mathscr{L}}{\delta
A_\mu}-\partial_\mu\big(\frac{\delta
\mathscr{L}}{\delta(\partial_\mu A_\nu)}\big)=0.
\end{eqnarray}
If the Lagrange's equation holds for any situation, both terms must
be zero,
\begin{eqnarray}\label{222}
\frac{\delta \mathscr{L}}{\delta A_\mu}=0,
\end{eqnarray}
\begin{eqnarray}\label{333}
&&\partial_\mu\big(\frac{\delta \mathscr{L}}{\delta(\partial_\mu
A_\nu)}\big)=0.
\end{eqnarray}
Then put the $\mathscr{L}$ into Eq. (\ref{222}) and Eq. (\ref{333}),
nonlinear Maxwell equations is obtained
\begin{eqnarray}
% \frac{\delta L}{\delta(\partial_\mu A_\nu)}=-F^{\mu\nu}+k_1xF^{\mu\nu}+k_2y\tilde{F}^{\mu\nu}\\
%\partial_\mu(-F^{\mu\nu}+k_1xF^{\mu\nu}+k_2y\tilde{F}^{\mu\nu})=0\\
&&\partial_t\mathbf{E}'=\nabla\times \mathbf{B}',\\
&&\partial_t\mathbf{B}=-\nabla\times \mathbf{E},\\
&&\nabla\cdot \mathbf{E}'=0,\\
&&\nabla\cdot \mathbf{B}=0,
\end{eqnarray}
where
\begin{eqnarray}
&&\mathbf{E}'=(1-k_1 x)\mathbf{E}-k_{2}y\mathbf{B},\\
&&\mathbf{B}'=(1-k_{1}x)\mathbf{B}+k_{2}y\mathbf{E},\\
&&x=-2(\mathbf{E}^2-\mathbf{B}^2),\\
&&y=-4\mathbf{E}\cdot\mathbf{B}.
\end{eqnarray}
Constructed with $\mathbf{E}$ and $\mathbf{B}$, $\mathbf{E}'$ and
$\mathbf{B}'$ take over their positions.

In order to get the Dirac form of nonlinear Maxwell equations, we
combine Eq. (14) and Eq. (15)
\begin{eqnarray}\label{444}
{ \partial_t}\begin{bmatrix}
 \mathbf{E}'\\
  i\mathbf{B}
\end{bmatrix}=\mathbf{\bigtriangledown}\times\begin{bmatrix}
\mathbf{B}'\\
-i\mathbf{E}\\
\end{bmatrix}.
\end{eqnarray}
Substituting expressions (20) and (21) into Eq. (\ref{444}), we
separate $\mathbf{E}$ and $\mathbf{B}$ after partial differential
operators $\partial_t$ and $\nabla$ act on $\mathbf{E}'$ and
$\mathbf{B}'$
\begin{eqnarray}\label{555}
&&M \partial_t\begin{bmatrix}
 E_{1}\\
  E_{2}\\
  E_{3}\\
  iB_{1}\\
iB_{2}\\
  iB_{3}
\end{bmatrix}=N \begin{bmatrix}
 iE_{1}\\
  iE_{2}\\
  iE_{3}\\
  B_{1}\\
B_{2}\\
  B_{3}
\end{bmatrix},
\end{eqnarray}
$M$ and $N$ are $6\times 6$ matrices involving  $\mathbf{E}$,
$\mathbf{B}$ and space partial differential operators. See Matrix
elements of $M$ and $N$ in Appendix.

In order to retain only time partial differential operator in the
left side, both sides of Eq. (\ref{555}) are multiplied by $M^{-1}$,
\begin{eqnarray}\label{999}
\partial_t\begin{bmatrix}
 E_{1}\\
  E_{2}\\
  E_{3}\\
 i B_{1}\\
iB_{2}\\
 i B_{3}
\end{bmatrix}=M^{-1}N \begin{bmatrix}
 iE_{1}\\
  iE_{2}\\
  iE_{3}\\
  B_{1}\\
B_{2}\\
  B_{3}
\end{bmatrix},
\end{eqnarray}
then $i\hbar$ times both sides of Eq. (\ref{999})
\begin{eqnarray}\nonumber
&& i\hbar\partial_t\begin{bmatrix}
 E_{1}\\
  E_{2}\\
  E_{3}\\
  iB_{1}\\
iB_{2}\\
 i B_{3}
\end{bmatrix}=i\hbar M^{-1}N \begin{bmatrix}
 iE_{1}\\
 i E_{2}\\
  iE_{3}\\
  B_{1}\\
B_{2}\\
  B_{3}
\end{bmatrix}\\
&&=-\hbar M^{-1}N\begin{bmatrix}
 1&0&0&0&0&0\\
 0&1&0&0&0&0\\
 0&0&1&0&0&0\\
  0&0&0&-1&0&0\\
0&0&0&0&-1&0\\
  0&0&0&0&0&-1
\end{bmatrix}\begin{bmatrix}
 E_{1}\\
 E_{2}\\
  E_{3}\\
  iB_{1}\\
iB_{2}\\
  iB_{3}
\end{bmatrix}.
\end{eqnarray}
If we assume
\begin{eqnarray}
H=-\hbar M^{-1}N\begin{bmatrix}
 1&0&0&0&0&0\\
 0&1&0&0&0&0\\
 0&0&1&0&0&0\\
  0&0&0&-1&0&0\\
0&0&0&0&-1&0\\
  0&0&0&0&0&-1
\end{bmatrix},
\end{eqnarray}
and
\begin{eqnarray}\label{666}
 \psi=\begin{bmatrix}
 E_{1}\\
  E_{2}\\
  E_{3}\\
  iB_{1}\\
iB_{2}\\
 i B_{3}
\end{bmatrix},
\end{eqnarray}
then,
\begin{eqnarray}\label{888}
i\hbar\partial_t\psi=H\psi.
\end{eqnarray}
Regard Eq. (\ref{888}) as the quantum equation of Maxwell equations
including virtual processes (a photon the transform into a
electron-positron pairs and interact with another photons).

We decompose $H$ into two parts: Hamiltonian $H_0$ of classical
electromagnetic fields and  Hamiltonian $H_{int}$ concluding virtual
particles
\begin{eqnarray}
H\psi=(H_0+H_{int})\psi.
\end{eqnarray}
$H_0$ is almost identical to Eq. (16), which exists when the wave
function has 6 components,
\begin{eqnarray}
H_0=-\hbar\begin{bmatrix}
 0 & 0  &  0  &  0  &\partial_3& -\partial_2\\
 0 & 0  &  0  &-\partial_3&  0  &\partial_1\\
 0 & 0  &  0  &\partial_2&-\partial_1& 0\\
 0 &\partial_3&-\partial_2&  0 & 0 & 0 \\
 -\partial_3& 0  &\partial_1& 0 & 0 & 0  \\
 \partial_2 &-\partial_1& 0  &0&0&0\\
\end{bmatrix}.
\end{eqnarray}
$H_0$ can be expressed with $\mathbf{s}$ and $\mathbf{p}$
\begin{eqnarray}
&&H_0=\begin{bmatrix}0 &  \mathbf{s}\cdot \mathbf{p}\\
\mathbf{s}\cdot\mathbf{ p} &0\end{bmatrix}.
\end{eqnarray}
$p_k$ is momentum operater and Eq. (1) has given the exactly form of
$s_k$. $H_{int}$ could also be written in a compact manner
\begin{eqnarray}
H_{int}=\Gamma^k\cdot p_k,
\end{eqnarray}
where $\Gamma^k$ are  $6\times6$ block matrices  and consist of
$3\times3$ matrixes $\mathcal {E}_k$ and $\mathcal {B}_k$ except
zero elements
\begin{eqnarray}
 \Gamma^k=i\begin{bmatrix}
-\mathcal {E}_k&\mathcal {B}_k
\\ 0&0\end{bmatrix}.
\end{eqnarray}
Elements of $\mathcal {E}_k$ and $\mathcal {B}_k$ are given in
appendix and they satisfy the following relations
\begin{eqnarray}
&&\mathcal {E}_k=\mathcal {E}_k^\texttt{T},\\
&&\mathcal{B}_k^{ij}(\mathbf{E},\mathbf{B})=-\mathcal{B}_k^{ji}(\mathbf{B},\mathbf{E}),
\end{eqnarray}
where $i,j,k$ run from 1 to 3. Eq. (37) means that, if we exchange
$\mathbf{E}$ and $\mathbf{B}$ in the two elements which are
symmetrical about  the diagonal of $\mathcal {H}_{int}$ and reverse
its sign, the two elements are equal.

In order to gain the same wave function as Eq. (6), transformation
is made
\begin{eqnarray}\label{777}
 i\hbar\partial_t U\begin{bmatrix}\mathbf{E}\\i\mathbf{B}\end{bmatrix}
=UH_0U^{-1}U\begin{bmatrix}\mathbf{E}\\i\mathbf{B}\end{bmatrix}+UH_{int}U^{-1}U\begin{bmatrix}\mathbf{E}\\i\mathbf{B}\end{bmatrix},
\end{eqnarray}
where $U$ is the transformation matrix
\begin{eqnarray}
&&U=\frac{1}{\sqrt{2}}\begin{bmatrix}1&1\\1&-1\\\end{bmatrix},
\end{eqnarray}
and $U^{-1}$ is the inverse matrix of $U$. Then Eq.(\ref{777}) can
be written
\begin{eqnarray}
i\hbar\partial_t
 \begin{bmatrix}\mathbf{E}+i\mathbf{B}\\\mathbf{E}-i\mathbf{B}\end{bmatrix}=\mathcal {H}_0\begin{bmatrix}\mathbf{E}+i\mathbf{B}\\\mathbf{E}-i\mathbf{B}\end{bmatrix}
 +\mathcal
 {H}_{int}\begin{bmatrix}\mathbf{E}+i\mathbf{B}\\\mathbf{E}-i\mathbf{B}\end{bmatrix},
\end{eqnarray}
where $\mathcal {H}_0$ and $\mathcal {H}_{int}$ is the new
Hamiltonian
\begin{eqnarray}
\mathcal {H}_0=UH_0U^{-1}=\begin{bmatrix}\mathbf{s}\cdot \mathbf{p} & 0\\0 &\mathbf{s}\cdot \mathbf{p}\\\end{bmatrix},\\
\mathcal {H}_{int}=\frac{i}{2}\begin{bmatrix}-\mathcal
{E}_k+\mathcal{B}_k &-\mathcal{E}_k-\mathcal{B}_k
\\-\mathcal {E}_k+\mathcal{B}_k&-\mathcal {E}_k-\mathcal{B}_k\\\end{bmatrix}\cdot
p_k.
\end{eqnarray}
Obviously, the linear part of the Hamiltonian $\mathcal {H}_0$,
which describes the classical electrodynamics, is a Hermitian
operator. However, $\mathcal {H}_{int}$ is not a Hermitian operator
although the relation between its elements shown in Eq. (36) and Eq.
(37) is enlightening. Based on the Hamiltonian of Heisenberg and
Euler which describes two photons scatter off one another in empty
space(virtual particles are considered), we have got the Hamiltonian
including $\mathcal {H}_{int}$. And the Hamiltonian is apparently
not Hermitian. Accordingly, the number of photons is not
conservative in the colliding process of two photons`.

\section{conclusion \label{conclu}}
In conclusion, starting from the Lagrangian $\mathscr{L}$ of quantum
electrodynamics in vacuum, we give the corresponding  Maxwell
equations. Then a new approach which is different from the tradional
method to derive from the form of nonlinear $Schr\ddot{o}dinger$
equation. At last, we transform the wave function in order to obtain
Riemann-Silberstein-Majorana-Oppenheimer wave vector representation.
\begin{acknowledgments}
This work is supported in part by NSF of China (Grants No.
10975075), Program for New Century Excellent Talents in University,
and the Project-sponsored by SRF for ROCS, SEM.
\end{acknowledgments}

\appendix
%\begin{appendices}
%\include{Appendix}
\section{Elements of $\mathcal {E}_k$ and $\mathcal {B}_k$ \label{app}}
\begin{eqnarray}
\mathcal {E}_1:
 &&\mathcal {E}_1^{11}=0,\\\nonumber
 &&\mathcal {E}_1^{22}=-8k_1E_2B_3+8k_2E_3B_2,\\\nonumber
 &&\mathcal {E}_1^{33}=8k_1E_3B_2-8k_2E_2B_3,\\\nonumber
  &&\mathcal {E}_1^{12}= \mathcal {E}_1^{21}=-4k_1E_1B_3+4k_2E_3B_1,\\\nonumber
 &&\mathcal {E}_1^{13}= \mathcal {E}_1^{31}=4k_1E_1B_2-4k_2E_2B_1,\\\nonumber
  &&\mathcal {E}_1^{23}=\mathcal {E}_1^{32}=4((k_1-k_2)(E_2B_2-E_3B_3).
\end{eqnarray}
\begin{eqnarray}
\mathcal {E}_2:
 &&\mathcal {E}_2^{11}=8k_1E_1B_3-8k_2E_3B_1,\\\nonumber
 &&\mathcal {E}_2^{22}=0,\\\nonumber
 &&\mathcal {E}_2^{33}=-8k_1E_3B_1-8k_2E_1B_3,\\\nonumber
 &&\mathcal {E}_2^{12}=\mathcal {E}_2^{21}=4k_1E_2B_3-4k_2E_3B_2,\\\nonumber
 &&\mathcal {E}_2^{13}= \mathcal {E}_2^{31}=4(k_1-k_2)(E_3B_3-E_1B_1),\\\nonumber
  &&\mathcal {E}_2^{23}=\mathcal {E}_2^{32}=-4k_1E_2B_1+4k_2E_1B_2.
\end{eqnarray}
\begin{eqnarray}
\mathcal {E}_3:
 &&\mathcal {E}_3^{11}=-8k_1E_1B_2+8k_2E_2B_1,\\\nonumber
 &&\mathcal {E}_3^{22}=8k_1E_2B_1-8k_2E_1B_2,\\\nonumber
 &&\mathcal {E}_3^{33}=0,\\\nonumber
 &&\mathcal {E}_3^{12}= \mathcal {E}_3^{21}=4(k_1-k_2)(E_1B_1-E_2B_2),\\\nonumber
 &&\mathcal {E}_3^{13}= \mathcal {E}_3^{31}=-4k_1E_3B_2+4k_2E_2B_3,\\\nonumber
  &&\mathcal {E}_2^{23}=\mathcal {E}_2^{32}=4k_1E_3B_1-4k_2E_1B_3.
\end{eqnarray}
\begin{eqnarray}
\mathcal{B}_1:
 &&\mathcal{B}_1^{11}=0,\\\nonumber
 &&\mathcal{B}_1^{22}=4(k_1-k_2)(B_2B_3-E_2E_3),\\\nonumber
 &&\mathcal{B}_1^{33}=4(k_1-k_2)(E_2E_3-B_2B_3)\\\nonumber
 &&\mathcal{B}_1^{12}=-4k_1E_1E_3-4k_2B_1B_3,\\\nonumber
 && \mathcal{B}_1^{21}=4k_1B_1B_3+4k_2E_3B_1,\\\nonumber
 &&\mathcal{B}_1^{13}=4k_1E_1E_2+4k_2B_1B_2,\\\nonumber
 &&\mathcal{B}_1^{31}=-4k_1B_1B_2-4k_2E_1E_2,\\\nonumber
  &&\mathcal{B}_1^{23}=4k_1(E_2^2+B_3^2)+4k_2(E_3^2+B_2^2),\\\nonumber
  &&\mathcal{B}_1^{32}=-4k_1(E_3^2+B_2^2)-4k_2(E_2^2+B_3^2).
\end{eqnarray}
\begin{eqnarray}
\mathcal{B}_2:
 &&\mathcal{B}_2^{11}=4(k_1-k_2)(E_1E_3-B_1B_3),\\\nonumber
 &&\mathcal{B}_2^{22}=0,\\\nonumber
 &&\mathcal{B}_2^{33}=-4(k_1-k_2)(E_1E_3-B_1B_3)\\\nonumber
 &&\mathcal{B}_2^{12}=-4k_1B_2B_3-4k_2E_2E_3,\\\nonumber
 &&\mathcal{B}_2^{21}=4k_1E_2E_3+4k_2B_2B_3,\\\nonumber
 &&\mathcal{B}_2^{13}=-4k_1(E_1^2+B_3^2)-4k_2(E_3^2+B_1^2),\\\nonumber
 &&\mathcal{B}_2^{31}=4k_1(E_3^2+B_1^2)+4k_2(E_1^2+B_3^2),\\\nonumber
  &&\mathcal{B}_2^{23}=-4k_1E_1E_2-4k_2B_1B_2,\\\nonumber
  &&\mathcal{B}_2^{32}=4k_1B_1B_2+4k_2E_1E_2.
\end{eqnarray}
\begin{eqnarray}
\mathcal{B}_3:
 &&\mathcal{B}_3^{11}=-4(k_1-k_2)(E_1E_2-B_1B_2),\\\nonumber
 &&\mathcal{B}_3^{22}=4(k_1-k_2)(E_1E_2-B_1B_2),\\\nonumber
 &&\mathcal{B}_3^{33}=0,\\\nonumber
 &&\mathcal{B}_3^{12}=4k_1(E_1^2+B_2^2)+4k_2(E_2^2+B_1^2),\\\nonumber
 && \mathcal{B}_3^{21}=-4k_1(E_2^2+B_1^2)-4k_2(E_1^2+B_2^2),\\\nonumber
 &&\mathcal{B}_2^{13}=4k_1B_2B_3+4k_2E_2E_3,\\\nonumber
 &&\mathcal{B}_2^{31}=-4k_1E_2E_3-4k_2B_2B_3,\\\nonumber
  &&\mathcal{B}_2^{23}=-4k_1B_1B_3-4k_2E_1E_3,\\\nonumber
  &&\mathcal{B}_2^{32}=4k_1E_1E_3+4k_2B_1B_3.
\end{eqnarray}
%\end{appendices}

\begin{thebibliography}{99}
\bibitem{Greiner1}W. Greiner, \emph{ Relativistic Quantum Mechanics},
(Springer-Verlag, Berlin, 1990), p105.
\bibitem{Dirac}
    P. Dirac, Proc. Roy. Soc. \textbf{117}, 610 (1928).
\bibitem{Landau}
    L. D. Landau and R. Peierls, Zeitschrift f$\ddot{u}$r Physik A Hadrons and Nuclei, \textbf{62}, 188 (1930).
\bibitem{Pauli}
    W. Pauli, Prinzipien der Quantentheorie, Handbuch der Physik, Vol. 24, Springer, Berlin, 1933, (English translation: General
      Principles of Quantum Mechanics, Springer, Berlin, 1980).
\bibitem{L&U}
    O. Laporte, G. E. Uhlenbeck, Phys. Rev. \textbf{37}, 1380 (1931).
\bibitem{Sil}
      L. Silberstein, Annalen der Physik, \textbf{327}, 579 (1906).
\bibitem{Opp}
    J. R. Oppenheimer,  Phys. Rev. \textbf{38}, 725 (1931).
\bibitem{1974}
    E. Mignani, E. Recami, M. Baldo, Lettere Al Nuovo Cimento, \textbf{11},
568 (1974).
\bibitem{Good}
      R. H. Good, Phys. Rev. \textbf{105}, 1914 (1957).
\bibitem{IBB1}
     I. Bialynicki-Birula, Acta physica Polonica. \textbf{86},
97 (1994).
\bibitem{I2}
     I. Bialynicki-Birula, Acta Physica Polonica B, \textbf{37},
935 (2006).
\bibitem{I3}
     I. Bialynicki-Birula, Phys Rev A, \textbf{79}, 032112 (2009).
\bibitem{Ger1}
     A. Gersten, Ann. Fond. L. de Broglie, \textbf{21}, 67 (1996).
\bibitem{Ger2}
    A. Gersten, Foundations of Physics Letters, \textbf{12},
291 (1999).
\bibitem{Ger3}
    A.Gersten, Foundations of Physics, \textbf{31},1211 (2001)
\bibitem{1956}
    T.Ohmura, Progress of Theoretical Physics, \textbf{16},
684 (1956).
\bibitem{19}
    R. Cook, Phys. Rev. A. \textbf{25}, 2164 (1982).
\bibitem{82}
    R. Cook, Phys. Rev. A. \textbf{26}, 2754 (1982).
\bibitem{1985}
    E. Giannetto, Lett. Nuovo Cim. \textbf{44},140 (1985).
\bibitem{1988}
    H. N$\ddot{u} \tilde{e}$z, A. Salas Brito, A. Salas Brito, C.
Vargas, Revista Mexicana de Fisica. \textbf{34}, 636 (1988).
\bibitem{1994}
    T. Inagaki, Phys. Rev. A. \textbf{49}, 2839 (1994).
\bibitem{1995}
    J. Sipe, Phys. Rev. A. \textbf{52}, 1875 (1995).
\bibitem{1998}
    S. Esposito, Found. Phys. \textbf{28}, 231 (1998).
\bibitem{I05}
     T. Ivezi$\acute{c}$, Found. Phys. \textbf{35}, 1585 (2005).
\bibitem{I052}
     T. Ivezi$\acute{c}$, Found. Phys. Lett. \textbf{18}, 401 (2005).
\bibitem{06}
     T. Ivezi$\acute{c}$, EJTP, \textbf{10}, 131 (2006).
\bibitem{2009}
     A. Bogush, V. Re$d^,$¡¯kov, N. Tokarevskaya, G. Spix, arXiv:0905.0261v1 (2009).
\bibitem{Greiner}
     W. Greiner, \emph{Quantum Electrodynamics},
(Springer-Verlag, Berlin, 2002), p426.
\end{thebibliography}
\end{document}